\documentclass[12pt,preprint]{aastex}
\usepackage{color}
\usepackage{graphicx}

\newcommand  \Hubble   {\ifmmode {\rm km\,s}^{-1}\,{\rm Mpc}^{-1}
                        \else km\,s$^{-1}$\,Mpc$^{-1}$\fi}
\newcommand  \Msun     {\ifmmode M_{\odot} \else M$_{\odot}$\fi}
\newcommand  \Lsun     {\ifmmode L_{\odot} \else $L_{\odot}$\fi}
\newcommand  \cms      {\ifmmode {\rm cm\,s}^{-1} \else cm\,s$^{-1}$\fi}
\newcommand  \acc      {\ifmmode {\rm km\,s}^{-2} \else km\,s$^{-2}$\fi}
\newcommand  \kms      {\ifmmode {\rm km\,s}^{-1} \else km\,s$^{-1}$\fi}

\newcommand  \ergs     {\ifmmode {\rm erg\,s}^{-1} \else erg s$^{-1}$\fi}
\newcommand  \ergcms   {\ifmmode {\rm erg\,cm}^{-2}\,{\rm s}^{-1}
                        \else erg\,cm$^{-2}$\,s$^{-1}$\fi}
\newcommand  \ergcmsA  {\ifmmode{\rm erg\,cm}^{-2}\,{\rm s}^{-1}\,{\rm \AA}^{-1}
                        \else erg\,cm$^{-2}$\,s$^{-1}$\,\AA$^{-1}$\fi}
\newcommand  \ergcmsHz {\ifmmode{\rm erg\,cm}^{-2}\,{\rm s}^{-1}\,{\rm Hz}^{-1}
                        \else erg\,cm$^{-2}$\,s$^{-1}$\,Hz$^{-1}$\fi}
\newcommand  \phcms    {\ifmmode {\rm photons\,cm}^{-2}\,{\rm s}^{-1}
                        \else photons\,cm$^{-2}$\,s$^{-1}$\fi}
\newcommand  \phcmsA   {\ifmmode {\rm photons\,cm}^{-2}\,{\rm s}^{-1}\,{\rm\AA}^{-1}
                        \else photons\,cm$^{-2}$\,s$^{-1}$\,\AA$^{-1}$\fi}
\newcommand  \rblr     {R$_{\rm BLR}$}

\begin{document}

\title{Broad Band Photometric Reverberation Mapping of NGC 4395}
\author{Haim Edri\altaffilmark{1}, Stephen E. Rafter\altaffilmark{1}, Doron Chelouche\altaffilmark{2}, Shai Kaspi\altaffilmark{1}, Ehud Behar\altaffilmark{1}}
\altaffiltext{1}{Physics Department, the Technion, Haifa 32000, Israel; e-mail: rafter, shai, behar: @physics.technion.ac.il}
\altaffiltext{2}{Physics Department, Faculty of Natural Sciences, University of Haifa, Israel; email: doron@sci.haifa.ac.il}

\begin{abstract}
We present results of broad band photometric reverberation mapping (RM) to measure the radius of the broad line region, and subsequently the black hole mass (M$_{\rm BH}$), in the nearby, low luminosity active galactic nuclei (AGN) NGC 4395.  Using the Wise Observatory's 1m telescope equipped with the SDSS g$'$, r$'$ and i$'$ broad band filters, we monitored NGC 4395 for 9 consecutive nights and obtained 3 light curves each with over 250 data points.  The g$'$ and r$'$ bands include time variable contributions from H$\beta$ and H$\alpha$ (respectively) plus continuum.  The i$'$ band is free of broad lines and covers exclusively continuum.  We show that by looking for a peak in the difference between the cross-correlation and the auto-correlation functions for all combinations of filters, we can get a reliable estimate of the time lag necessary to compute M$_{\rm BH}$.  We measure the time lag for H$\alpha$ to be $3.6 \pm 0.8 $ hours, comparable to previous studies using the line resolved spectroscopic RM method.  We argue that this lag implies a black hole mass of M$_{\rm BH} = (4.9 \pm 2.6) \times 10^{4}$ \Msun .
\end{abstract}

\keywords{galaxies: Seyfert --- galaxies: individual (NGC 4395) --- techniques: photometric reverberation mapping}

\section{Introduction}
Active galactic nuclei (AGN) in general are known to be composed of a super massive black hole (BH), whose mass ranges from $10^{5}\--10^{9}$ \Msun, and is surrounded by a disk of accreting material.  The accretion disk is the source of optical and ultraviolet (UV) continuum emission.  Farther out, but still within the gravitational influence of the BH are high velocity (FWHM $>800 \kms$) clouds in assumed Keplerian orbits.  This region is referred to as the broad line region (BLR) due to the Doppler broadening (FWHM $\approx 800\--10,000$ \kms) of the permitted emission lines produced there.  H and He produce the prominent emission lines in the optical, while C, N and O, as well as the H and He Lyman lines are prominent in the UV.

An important characteristic of AGNs is the time variability of the continuum and the broad emission lines.  The classic spectroscopic reverberation mapping (RM) technique is based on the delayed response of the emission lines produced in the BLR to the variable central continuum source.  The time lag, $\tau$, is due to the light travel time from the central source to the BLR and is thus proportional to the BLR distance (\rblr\ $=\rm c\tau$).  By combining the velocity of the BLR clouds (assuming Keplerian orbits) along with their distance we can determine the total mass contained within the BLR (which is dominated by the BH) in a simple way using

\begin{equation}
{\rm M_{\rm BH}}=\frac{f{\Delta V}^2 c \tau}{G}, 
\end{equation}
where G is the gravitational constant and $f$ is a factor (of order unity and discussed below) that depends on the geometric and kinematic structure of the BLR \citep[e.g.,][]{2000ApJ...533..631K}.

Until now, RM was mostly done using resolved line spectroscopy \citep[see review by][]{1993PASP..105..247P}.  Although this method allows one to accurately measure the flux in each line, it is observationally time consuming.  Only about 45 objects have been mapped using this method in the last two decades \citep{2000ApJ...533..631K, 2005ApJ...619L.151B, 2010ApJ...721..715D, 2009ApJ...697..160B}. Recently, narrow band photometric RM was tested \citep{2011A&A...535A..73H}, but the narrow bands must be fitted to the redshift of the observed objects and their emission lines.  \citet{2012ApJ...747...62C} suggested an efficient approach to photometric RM which offers a relatively quick, although somewhat less sensitive method, using broad band images instead of spectra or narrow band filters.  Upcoming photometric surveys such as the Large Synoptic Survey Telescope (LSST) are planned to continuously monitor at least $10^{7}$ quasars ($0 < z < 6$) over the next decade \citep{2007AAS...21113702I}.  Broad band photometric RM can utilize such data to estimate the mass of numerous quasars and increase by several orders of magnitude the number of reverberation mapped objects.  It is particularly useful in cases, such as the present one, where spectroscopic integration times for sufficient signal-to-noise (S/N) are too long compared to the lag. 

The purpose of this study is to test the broad band photometric RM method \citep{2012ApJ...747...62C} by measuring the time lag between the continuum and the broad line emission in an AGN.  We chose to monitor NGC 4395 since its BLR lag has been well studied in the past, and also because it is the least luminous Seyfert 1 galaxy known to date with $\lambda L_{\lambda}(\rm 5100\AA)\simeq 7 \times 10^{39}$ $\ergs $ and apparent g$'$ magnitude of 14.55 mag \citep{2009ApJS..182..543A}. The redshift of $z=0.001064$ implies for NGC 4395 a distance of 4.42 Mpc (using H$_{\rm 0} = \rm 72$ km s$^{-1}$ Mpc$^{-1}$).  Extrapolating  the relation between the AGN luminosity and the distance to the BLR \citep{2000ApJ...533..631K} to much lower luminosities, we expect a time lag of only a few hours.  A spectroscopic RM campaign in the UV with Hubble Space Telescope found a time lag of $\sim 1.0$ hour for the C{\sc iv} line \citep{2005ApJ...632..799P}. \citet[][hereafter D06]{2006ApJ...650...88D}, using the H$\alpha$ line, found a time lag of $\sim1.3$ hours.  Both of these papers use their lag measurements to infer a mass M$_{\rm BH}\approx 3.0 \times 10^{5}$ \Msun .

\section{Observations and Data Reduction}
NGC 4395 was observed with the Wise Observatory's 1 meter telescope that is equipped with the Sloan Digital Sky Survey (SDSS) g$'$, r$'$ and i$'$ filters, and a 1300 $\times$ 1340 PI (Princeton Instruments) CCD.  Each pixel corresponds to 0.58$''$ on the sky and the overall field of view is $12^{\prime}.6 \times 13^{\prime}.0$.  The g$'$ filter covers  the wavelengths 4000 \AA\ to 5500 \AA , r$'$ covers 5600 \AA\ to 7100 \AA , and i$'$ covers 7000 \AA\ to 8500 \AA\ (see Figure 1 for the SDSS filter transmission curves). We observed NGC 4395 for 9 nights during 2011, from February 23 to March 6, with an exposure time of 5 minutes per image.  This gives about 30 exposures per night, per filter, resulting in a total of about 250 images per filter.

Data reduction was performed using standard IRAF routines. Flux measurements were carried out using PSF photometry and the light curves for NGC 4395 in each of the three bands were produced by comparing its instrumental magnitudes to those of constant-flux stars in the field \citep[see, e.g.,][for details]{1996MNRAS.279..429N}. The uncertainties on the photometric measurements include the fluctuations due to photon statistics and the scatter in the measurement of the stars used. In order to calibrate the instrumental response in the SDSS g$'$, r$'$, and i$'$ bands, we used the SDSS catalog to find the absolute magnitude of the constant-flux stars in the field and used these to calculate the absolute magnitude of the NGC 4395 measurements. The light curves for the three filters are presented in Figure 2 and the data are listed in Table 1. The data points are clustered in nine nights.  The right-hand panel of Figure 2 shows the light curves from the first night.  Upon visual inspection, no obvious lag is apparent since the signal in all three bands is dominated by the continuum.  

Table 2 shows the variability parameters in each of the three filters.  Column (1) indicates the filter, column (2) shows the number of points in each light curve, column (3) is the minimum magnitude (M$_{\rm min}$), column (4) is the maximum magnitude (M$_{\rm max}$), column (5) is the mean magnitude (M$_{\rm mean}$), column (6) is the ratio of the maximum to minimum flux (R$_{\rm max}$).  Column (7) is the fractional flux variation defined as $F_{var}=\sqrt{\sigma^2-\Delta^2}/<F>$, where $\sigma$ is the standard deviation of the full light curve and $\Delta$ is the root mean square of the errors of the individual data points \citep[for details see][]{1997ApJS..110....9R,2002ApJ...568..610E}.  In order to calculate R$_{\rm max}$ and F$_{\rm var}$ we converted the light curves, which are given in magnitude (Figure 2), into flux using the conversion F $= 10^{(m_{0}-m)/2.5)}$ where m$_{0}$(g$'$) $=$ 25.11 mag, m$_{0}$(r$'$) $=$ 24.80 mag, and m$_{0}$(i$'$) $=$ 24.36 mag.  D06 find that F$_{\rm var}$ of the H$\alpha$ light curve measured spectroscopically is comparable to, or can even exceed F$_{\rm var}$ of the 5100 \AA\ continuum (see Table 4 of D06), and both are within the range of $\sim$ 0.03 $-$ 0.06, exactly the presently measured broad band F$_{\rm var}$ range shown in Table 2.  While NGC 4395 is an extremely low luminosity AGN, higher luminosity quasars such as those presented in \citet{2007ApJ...659..997K}, have F$_{\rm var}$ that range from 0.027$-$0.293.  Given the definition of F$_{\rm var}$, if $\sigma/\Delta \approx 1$ then F$_{\rm var}$ approaches zero and the ability to trace the lag needs to be tested.  Here the average ratio for the three light curves is $\sigma/\Delta = 2.3$.  In Section 3.2 we demonstrate that under these conditions a time lag can be accurately recovered.

\section{Analysis Method}
Instead of using the resolved lines of AGN, following \citet{2012ApJ...747...62C}, we use the broad band flux in the following way: the flux $f$ through a filter $X$ of the continuum is denoted by $f_{X}$. We assume here that filter $X$ has contribution from the continuum alone.  A measured flux through a second filter $Y$ that has both continuum and line contributions is denoted by $f_{Y}$. The total flux through filter Y as a function of time (i.e., the Y filter light curve) is

\begin{equation}
f_{Y}(t)=f_{Y}^{c}(t)+f_{Y}^{l}(t).
\end{equation} 
To measure the lag $\tau$ between the continuum ($f_{Y}^{c}$) and the line ($f_{Y}^{l}$), we need to compute the cross-correlation function (CCF) between these two components of the light curve

\begin{equation}
CCF(\tau)=f_{Y}^{l}(t + \tau)*f_{Y}^{c}(t)=(f_{Y}(t + \tau) - f_{Y}^{c}(t + \tau))*f_{Y}^{c}(t)
\end{equation}
where * denotes an integral over time (i.e., the convolution between the two functions).  In spectroscopic RM the peak of the CCF($\tau$) gives the required time lag.  We now assume the time variability of the continuum flux in the Y band is the same as that in the X band which contains exclusively continuum, namely $f_{Y}^{c}(t)\approx f_{X}(t)$.  This is a good approximation in the optical since the continuum is 75\%\ - 95\%\ of the total flux in the filters, where the remaining variable flux is mostly from the broad lines.  We can now write

\begin{equation}
CCF(\tau)\approx (f_{Y}(t + \tau)-f_{X}(t + \tau))*f_{X}(t) \approx CCF_{XY}(\tau)-ACF_{X}(\tau)
\label{ACF}
\end{equation} 
\noindent where ACF is the auto$-$correlation function. Equation 4 is the crux of this method, namely the difference between the CCF and the ACF being used to determine $\tau$ and not just the CCF alone. Essentially, by having broad band photometric light curves, we can build the line-continuum cross correlation function and estimate the time lag.\\

In order to detect a time lag between the lines and the continuum in NGC 4395, two filters were chosen which cover the H$\alpha$ and H$\beta$ lines (i.e., $f_{Y}$ defined above).  A third filter was chosen to cover a continuum band that is relatively free of lines (i.e., $f_{X}$).  This is demonstrated in Figure 1 using the SDSS g$'$, r$'$ and i$'$ filters for NGC 4395, and would be similar for objects with similarly low redshifts.  For objects at higher redshift where  Balmer lines continually shift to longer wavelengths, a careful choice of filters is required to apply the method.  In NGC 4395, while the Balmer decrement is about 4, the broad component of the H$\alpha$ line contributes 7\% to the flux in the r$'$ band, while H$\beta$ contributes only about 3\% to the flux in the g$'$ band.  Consequently, the i$'$ and g$'$ bands are dominated by the continuum, while the r$'$ band stands the best chance to reveal the BLR variability.  Nonetheless, we computed CCFs and ACFs for all filter combinations.  The difference between the CCF and the ACF gives the sought after function, of which the peak determines the time lag between the line and the continuum.  The results for the three filter pairs are presented in section \ref{results}.

\subsection{Time Series Analysis}
A complication to consider when calculating the CCF and ACF is the non-uniform sampling of astronomical data.  In order to overcome this difficulty we use the interpolated cross-correlation function method \citep[ICCF,][]{1998PASP..110..660P}.  Following \cite{1999PASP..111.1347W}  we use the local ICCF method. In this method the mean and standard deviation are calculated at every time step, taking into account only the values within the overlapping parts of the light curves, giving more reliable time lags.

Since the present monitoring is rather uniform throughout a given night (15 minute intervals), and in order to avoid unnecessary interpolation that adds artificial data to the light curves, we used an algorithm which is slightly different than the method described by \cite{1998PASP..110..660P}.  While cross-correlating two functions, one function is shifted against the other in fixed time steps.  We choose this time step to be determined by the original data, i.e we allow the ICCF to be unevenly sampled along the $\tau$ axis.  This approach enables gaps in the data (e.g., intranight) to be skipped, thereby minimizing unnecessary interpolation.  The method was tested by simulating light curves made from unevenly sampled data. The results of the simulations together with the error estimates are explained in the next section and confirm that the correct time lag is recovered in each simulation. 

The final product we are seeking is the difference between the CCF($\tau$) and ACF($\tau$) (Equation \ref{ACF}).  Since the time lags ($\tau$) for which the two functions are calculated are not exactly the same, we subtract the closest data points of the two functions (i.e., no interpolation).  However, since the difference in sampling is small (a few minutes), this is a very good approximation.  In order to calculate the time lag from the CCF($\tau$)-ACF($\tau$) difference we first identify the maximum.  This peak represents the lag.  Subsequently, we retain only the region around the maximum that is greater than 80\% of the peak and from this interval the time lag centroid is found by a simple weighted average calculation \citep{1998PASP..110..660P}.

\subsection{CCF Simulations and Uncertainties}
In order to estimate the time lag and its uncertainty we need to propagate the magnitude errors in the light curves to an error for the time lag.  Following \cite{1998PASP..110..660P},  the flux randomization (FR) and random subset selection (RSS) methods are used for this in conjunction with Monte Carlo (MC) simulations. For the FR, one alters the magnitude measurement within the errors. Each data point is modified according to a Gaussian distribution around the measured value, with a width that is the measurement uncertainty. The modification of each data point is statistically independent of the other data points.  The RSS is used in order to estimate the error due to unevenly sampled data by randomly excluding data points from the simulated light curves. Thus, each realization is based on a randomly chosen subset of the original data points.  The FR procedure tests the sensitivity of the results to the flux accuracy of the measurement, while the RSS method checks the effect of the incomplete sampling.  In a single MC realization which includes both the FR and the RSS simulations, one generates two light curves, then the CCF($\tau$)-ACF($\tau$) difference is computed and the lag is recorded. The resulting distribution of 10,000 realizations is subsequently analyzed.

The distribution of realizations can be used to estimate both the expected lag from the centroid of the distribution, and the statistical error from the width.  For each pair of light curves the lag was calculated using the FR method only and the FR/RSS combined method.  These results are also shown in Figure 3.  While the FR method gives the correct time lags with errors of about $\pm$1 hour, the FR/RSS method shows not only larger errors, but also overestimates the time lags by up to 50\%. Using the FR/RSS method on the actual NGC 4395 data resulted in no lag at all for two out of the three possible filter pairs.  We conclude that the RSS estimate is less appropriate here because of the highly regular, intranight sampling.  For each filter 4 images were taken within 1 hour, giving about 32 evenly sampled data points during each night.  For most nights the data were taken without significant interruption and since our method ignores the gaps between the nights there is no need to invoke the RSS method.  In fact, it can lead to erroneous results and exaggerated error estimates (see Figure 3).  Ignoring the RSS estimate is valid as long as the time lag is longer than both the variability and sampling time scales, which applies in the case of NGC 4395.

We wish to test whether one can recover a time lag given the low variability of the observed light curves (F$_{\rm var} = 0.03 - 0.06$).  For this purpose, the photometric i$'$ band light curve is taken as the `continuum' light curve.  In order to simulate the H$\alpha$ contribution, this `continuum' light curve is scaled down to 7\% of the continuum (i.e., the approximate variable broad H$\alpha$ contribution to the r$'$ band), uniformly shifted by an arbitrary lag of 4 hours, and added to the `continuum' light curve to simulate the photometric `line$+$continuum' light curve presented in Figure 4 (in Panel a we show only one night, but the simulation includes all 9 nights).  We then employ the above CCF method to test whether the input lag can be recovered.  As can be seen in Figure 4 (Panel c) the lag is easily recovered.  Subsequently, we simulated 10,000 light curves using the FR method and measured the lag on each one.  The distribution of lags are presented in Figure 4 (Panel d) and are centered around the correct lag, namely $4.0^{+1.5}_{-1.1}$.  The small peak near zero is a numerical artifact due to binning when computing the CCF and ACF.  

We also checked whether we can recover the input lag for an even less variable line contribution by repeating these simulations with (`line') F$_{\rm var}$ values as low as 0.01.  In all cases we could still recover the input time lag as long as the line contribution to the band remains at the level of a few percent; the main reason being the long duration, well cadenced sampling of the light curve over 9 full nights.  Clearly, it is the high statistics of the light curves that make the detection of the lag possible.  While trying to recover the lag from smaller and smaller sub-samples, one would expect the lag signal to weaken, down to the point where it can not be constrained to within meaningful errors.  Indeed, the lag value of 2.8 $-$ 4.4 hours does not change much when applying the method to sub samples, and as small as a third of the full light curve. Below this limit, the residual data are too poor to constrain the lag.  Evidently, some nights have a somewhat stronger lag signal than the others.

\section{Results \& Discussion} \label{results}
\subsection{Time Lag}
In Figure 5 we show the ACFs of all three filters.  Both the g$'$ and i$'$ bands show very similar ACFs, meaning that their light curves are very similar and dominated by variable continuum emission.  The r$'$ band, however, shows a departure in the ACF that is indicative of variations of both the continuum and an additional H$\alpha$ line emission component.  This difference in the r$'$ ACF is clearly seen at time lag of around 4 hours.  

Deriving the time lags from the CCF-ACF differences presented in Figure 6, we find the time lag between H$\alpha$ and the continuum to be $3.46^{+1.59}_{-0.36}$ hours using the r$'$ and the i$'$ filters (top panel).  The peak from which we measure this lag is similar to results of simulations as shown in \citet{2012ApJ...747...62C}.  Further, from a much clearer peak, we find a lag of $3.68^{+0.70}_{-0.84}$ hours using r$'$ and g$'$ (middle panel).  This also represents a lag between the continuum and H$\alpha$ since the g$'$ filter includes only a 3\% contribution from the H$\beta$ line, which is apparently negligible for the CCF analysis.  The errors quoted above are taken from the distributions shown in Figure 7 using the FR method for these filter combinations. The time lag between H$\beta$ and the continuum, using g$'$ and i$'$ (bottom panel), cannot be detected due to the weak contribution from the H$\beta$ line mentioned above.  Even with spectroscopic RM the lag of H$\beta$ to the continuum could not be detected (D06).  The top and bottom panels of Figure 6 show a negative time lag near 4 hours, which could indicate an emission component in the i$'$ band that lags the r$'$ and g$'$ bands.  On the other hand, there is no indication for this lag in the ACF profiles around 4 hours (Figure 5).  Hence, the viability and nature of the additional emission component in the i$'$ band is currently unclear.  The two positive time lags we do measure for the r$'$ band are consistent with each other, and averaging them gives a time lag between H$\alpha$ and the continuum of $3.6\pm 0.8$ hours.

The time lag of the BLR in NGC 4395 was estimated before by spectroscopic RM.  Comparison of the current H$\alpha$ lag with that reported in D06 is presented in Figure 8.  The blue circles denote the standard ICCF, and gray error bars with no symbols are the ZDCF  from D06.  Also shown is the CCF-ACF difference from this study, scaled up by a constant factor for comparison between the peaks.  The D06 time lag centroid (using the 80\% threshold criterion) is 2.6 hours, which is slightly less, but still comparable to the present CCF-ACF result.  The ZDCF has a broad peak at about 4 hours and also a secondary peak at about 7.5 hours.  This second peak we disregard given that each night of observations is only about 8 hours long.  As is often done in RM analyses \citep[see e.g.,][]{2000ApJ...533..631K}, the ZDCF results are included here to ensure that the main result (i.e., the ICCF peak) is not an artifact of the interpolation scheme used.  In fact, it is often the case that the ZDCF is more erratic than the ICCF, and may show a multi-peak structure (e.g., H$\alpha$ for PG0052, H$\alpha$ \& H$\beta$ in PG1226, H$\alpha$ in PG1211, H$\alpha$a \& H$\beta$ in PG1307 in \citet{2000ApJ...533..631K}).  Specifically, in the case of NGC4395, it is difficult to identify a turnover at $>$ 7 $-$ 8 hours since the typical night duration is of the same order. As such, it is not clear that a second peak exists, and more data beyond the typical 8 hours of observation per night are required to verify its significance.

\subsection{Black Hole Mass} 
While the main result of this work is the measured time lag, it would be useful to determine M$_{\rm BH}$ based on the current measurement in order to compare with those of previous works.  This requires a slight digression into the velocity of the BLR and the appropriate values to use in Equation 1.  There are still many open questions about which is the proper way to determine the `true' velocity of BLR clouds being mapped. The two methods used in classic spectroscopic RM determine the width of the line from either an RMS spectrum or from a mean spectrum.  For the broad band photometric RM method in particular, there is no RMS or mean spectrum so this is not an option.  Therefore, we must choose how to most appropriately estimate the H$\alpha$ velocity width.  To get some idea of the velocity and hence the mass, we must use a single spectrum which we assume, as in the single-epoch mass determination method \citep[see][and references therein]{2009ApJ...692..246D}, gives a reasonable measure of the velocity.

At this point, it is also debated whether one should use the FWHM or $\sigma$ of the emission line.  \citet{2011Natur.470..366K} show that as broad lines become more narrow (FWHM $< 4000$ km s$^{-1}$) their profiles become less Gaussian and more Lorentzian, and $\sigma$ becomes very large and thus, may not represent the `true' BLR velocity.  In this case, as is the case with NGC 4395, the FWHM becomes a better parametrization of the `true' velocity.  The scaling factor $f$ is also a debated topic for M$_{\rm BH}$ determination.  When scaling RM-based masses to the empirically determined M-$\sigma_{*}$ relation for quiescent galaxies, \citet{2010ApJ...716..269W} recently found that $f = 5.3$ while \citet{2011MNRAS.412.2211G} find a value that is roughly half ($f = 2.8$).  These studies are also limited by the lack of data on the very low mass end, where NGC 4395 is located, and it is still debated as to whether these low mass objects follow the same scaling relation \citep{2011arXiv1102.0537M, 2011ApJ...739...28X, 2011ApJ...741...66R}. \citet{2006A&A...456...75C}, in a study of RM AGNs, find a range in the value of $f$ based on the FWHM/$\sigma$ ratio.  For objects where this ratio is between 1.4 and 2.8 (as is the case for NGC 4395), $f$ ranges from 2.5 down to 0.71.  Indeed, from the simplest assumptions of the BLR as an isotropic circular velocity field where the FWHM represents twice the typical BLR velocity, one expects $f = 0.75$ \citep{2000ApJ...533..631K}.  Because the simplifying assumptions made in the RM case are already based on Keplerian orbits anyway, and since measuring $\tau$ in RM does not take into account kinematics, specific geometries or other effects (like radiation pressure for example) the most appropriate factors to use in Equation 1 are $f = 0.75$ along with the FWHM, which we use below to calculate M$_{\rm BH}$.

NGC 4395 has been observed spectroscopically many times over the past 20 years with several different instruments and wavebands.  The first high resolution spectra were obtained with the Keck telescope \citep[see][and references therein]{1999ApJ...520..564K} where the H$\beta$ FWHM is reported to be $\sim$ 1500 km s$^{-1}$.  However in the work of D06, the velocity used to compute M$_{\rm BH}$ was taken from the variable part of the H$\alpha$ line (i.e., the RMS spectrum found from their RM campaign) and is (FWHM $\approx \sigma$) $\sim$ 2300 km s$^{-1}$.  Since the velocity is squared in Equation 1 this leads to more than a factor of 2 difference in M$_{\rm BH}$.  It has also been suggested that there is a non-negligible intermediate line region in NGC 4395 that could make up a significant fraction of the total Balmer line profiles, but does not vary on these short time scales (A. Laor, private communication).  From a more recent, high resolution Keck spectrum \citep{2006ApJ...636...83L} the narrow lines with asymmetric blue wings are roughly FWHM $= 70$ km s$^{-1}$.  
 
A loosely constrained 3-component Gaussian fit to H$\beta$ in the Keck spectrum, which takes into account an intermediate velocity, yields a broad FWHM of $\sim$ 1175 $\pm$ 325 km s$^{-1}$, in addition to a intermediate width of 270 km s$^{-1}$, and a narrow line width of 70 km s$^{-1}$ constrained by the [O{\sc {iii}}] lines.  This broad component is roughly consistent with the measurement in \citet{1999ApJ...520..564K}.  Since we measure a time lag for the H$\alpha$ line it would be best to use the width of the same line for our mass determination.  However, it has been shown in \citet{2005ApJ...630..122G} for moderate luminosity Seyfert galaxies that H$\beta$ and H$\alpha$ have similar widths, on average to within $\sim$ 0.2 dex.  All things considered, we take a reasonable value of 1500 $\pm 500$ km s$^{-1}$ for the H$\alpha$ line to estimate M$_{\rm BH}$.  

Combining these results with Equation 1 we find that M$_{\rm BH} = (4.9 \pm 2.6) \times 10^{4}$ \Msun .  In Table 3 we compare the present results with those of \citet{2005ApJ...632..799P} and \citet{2006ApJ...650...88D}.  Note that \citet{2005ApJ...632..799P} would get a mass similar to the present mass estimate if they had used $f =$ 0.75 instead of $f= $5.5 using the C{\sc iv} lag and velocity.  The discrepancy between masses is exactly a factor of 7.3 (note that 5.5$/$0.75 $=$ 7.3).  Also note that the time lag quoted from D06 is lower than the centroid value derived from Figure 8 since the centroid is determined using the entire CCF, whereas we use the aforementioned 80\% criterion.

\section{Conclusions}
Broad band photometric RM is used here for the first time to measure the distance to the BLR in an AGN and to estimate M$_{ \rm BH}$.  It is shown to give a good estimate of the time lag between the variable continuum source and the response in the BLR of NGC 4395 when compared with spectroscopic RM results.  A clear time lag was determined for the H$\alpha$ line and the continuum to be $3.6\pm 0.8$ hours, but a lag between H$\beta$ and the continuum could not be detected, as expected.  This result is consistent with previous works on NGC 4395 (D06).  In AGNs like NGC 4395, where the H$\beta$ line is weak by comparison with the total integrated flux in a given filter, time lags will be difficult to measure, which is a limitation of the photometric RM method for such weak line.  Estimation of M$_{ \rm BH}$ can be made only by using the velocity of the BLR from a separate spectroscopic measurement.  We find for NGC 4395 M$_{ \rm BH}= (4.9 \pm 2.6) \times 10^{4}$ \Msun, which is significantly lower than previous estimates.  This is not due to the lag but due to the much smaller geometrical factor $f$ used in Equation 1, and is consistent with M$_{ \rm BH}$ derived from C{\sc iv}.

The broad band photometric RM method has the advantage that it is less time intensive observationally when compared to spectroscopic RM, while still yielding reliable results.  It is therefore particularly useful when applied to low luminosity AGNs where integration times on available spectrometers are comparable to the BLR time lag.  Photometric RM can be applied to future large scale surveys that will photometrically monitor numerous AGNs, potentially allowing M$_{\rm BH}$ to be determined for a large number of AGNs.  It is especially applicable to faint objects that cannot be monitored spectroscopically due to prohibitively long integration times or too low S/N.

\acknowledgments
We acknowledge helpful discussions with Ari Laor for providing useful insights into the nature of broad line profiles in AGNs.  This work is supported by a grant from the Israel Science Foundation.  S. R. is supported at the Technion by the Zeff Fellowship.  The research of D.C. is partially supported by ISF grant 927/11.  Based on observations obtained with the Tel Aviv University Wise Observatory 1 meter telescope.

\begin{deluxetable}{cccccc}
\tablecolumns{6}
\tabletypesize{\footnotesize}
\tablewidth{0pc}
\tablecaption{Photometric light curves for NGC 4395 in magnitude units.}
\tablehead{
	\colhead{JD-2455600} &
	\colhead{i$'$} & 
	\colhead{JD-2455600} &
	\colhead{r$'$} &
	\colhead{JD-2455600} &  
	\colhead{g$'$} \\
\colhead{(1)} & \colhead{(2)} & \colhead{(3)} & \colhead{(4)} & \colhead{(5)} & \colhead{(6)}
}
\startdata
16.28832 & $16.878 \pm 0.048$ & 16.29479 & $16.272 \pm 0.027$ & 16.30125 & $17.289 \pm 0.019$ \\
16.29803 & $16.928 \pm 0.035$ & 16.31416 & $16.270 \pm 0.026$ & 16.31093 & $17.264 \pm 0.022$ \\
16.30771 & $16.844 \pm 0.042$ & 16.32384 & $16.263 \pm 0.022$ & 16.32061 & $17.274 \pm 0.015$ \\
\enddata
\tablecomments{Table 1 is presented in its entirety online in the electronic version.}
\end{deluxetable}

\begin{deluxetable}{cccccccc}
\tablecolumns{7}
\tabletypesize{\footnotesize}
\tablewidth{0pc}
\tablecaption{Variability Parameters. }
\tablehead{
	\colhead{Time Series} &
	\colhead{Number of Data Points} & 
    \colhead{$M_{\rm min}^{a}$} & 
    \colhead{$M_{\rm max}^{a}$} & 
    \colhead{$M_{\rm mean}^{a}$} & 	
	\colhead{$R_{\rm max}^{b}$} & 
	\colhead{$F_{\rm var}^{b}$} \\
\colhead{(1)} & \colhead{(2)} & \colhead{(3)} & \colhead{(4)} & \colhead{(5)} & \colhead{(6)} & \colhead{(7)}
}
\startdata
g$'$ (Contains H$\beta$)  & 270 & 17.14 & 17.37 & 17.25 & 1.228 & 0.041 $\pm$ 0.002 \\
r$'$ (Contains H$\alpha$) & 282 & 16.15 & 16.35 & 16.25 & 1.207 & 0.037 $\pm$ 0.002 \\
i$'$ (Continuum)          & 271 & 16.72 & 17.01 & 16.87 & 1.307 & 0.054 $\pm$ 0.003 \\
\enddata
\tablecomments{a-magnitude, b-calculated from flux}
\end{deluxetable}

\begin{deluxetable}{cccc}
\tablecolumns{4}
\tabletypesize{\footnotesize}
\tablewidth{0pc}
\tablecaption{NGC 4395 Black Hole Mass Estimates}
\tablehead{
	\colhead{Reference} &
	\colhead{\cite{2005ApJ...632..799P}} & 
	\colhead{\cite{2006ApJ...650...88D}} & 
	\colhead{This study} \\
    \colhead{(1)} &	
    \colhead{(2)} &
 	\colhead{(3)} &
 	\colhead{(4)} 
 }
\startdata
Spectral Line			       & C{\sc iv}         & H$\alpha$         & H$\alpha$       \\
Time Lag (hours)                       & 0.95 $\pm$ 0.36   & 1.33 $\pm$ 1.13   & 3.6 $\pm$ 0.8   \\ 
Velocity (\kms, $\sigma \approx$ FWHM) & 2900 $\pm$ 300    & 2300              & 1500 $\pm$ 500  \\ 
$f$                                    & 5.5               & 5.5               & 0.75            \\ 
Mass ($10^{5}$ \Msun)                  & 3.6 $\pm$ 1.1     & $\sim$ 3.0        & 0.49 $\pm$ 0.26 \\ 
\enddata
\end{deluxetable}

\begin{figure}[t]
  \hskip -1.0cm
    \includegraphics[scale=0.5]{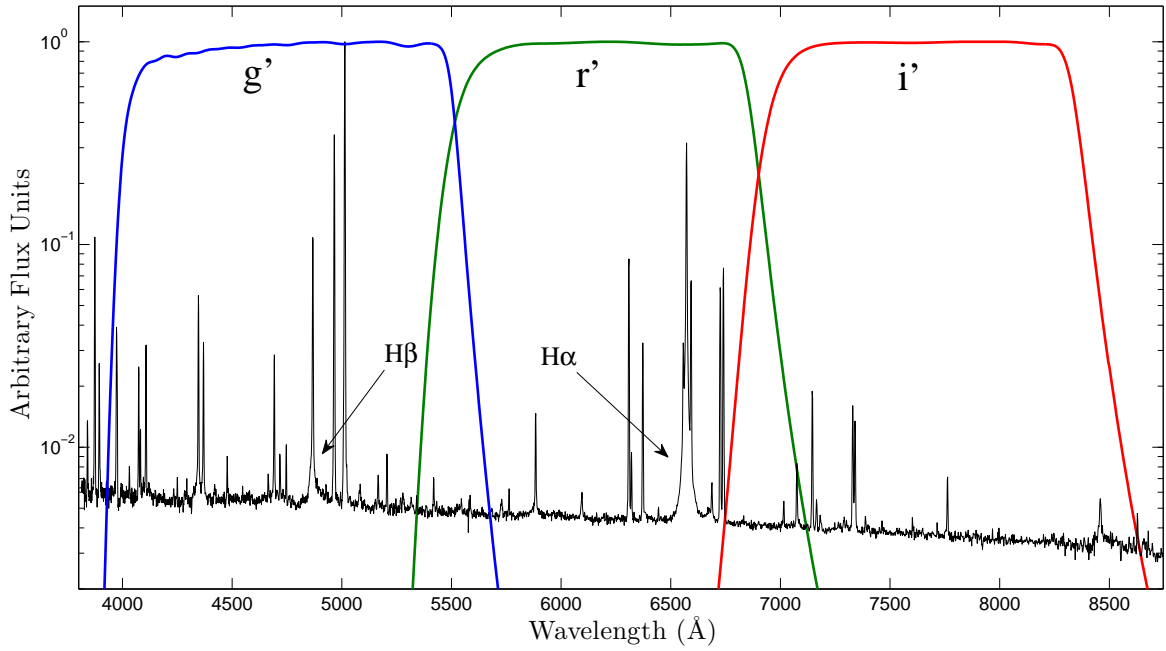}
    \caption{Observed spectrum of NGC 4395 from SDSS DR7. The H$\alpha$ line is at 6569 \AA\ and the H$\beta$ line is at 4866 \AA . The 3 filters we use are g$'$ that includes H$\alpha$, r$'$ that includes H$\beta$, and i$'$ that includes mostly only continuum.}
\end{figure}

\begin{figure}[]
  \hskip -10.0cm
    \includegraphics[scale=0.7]{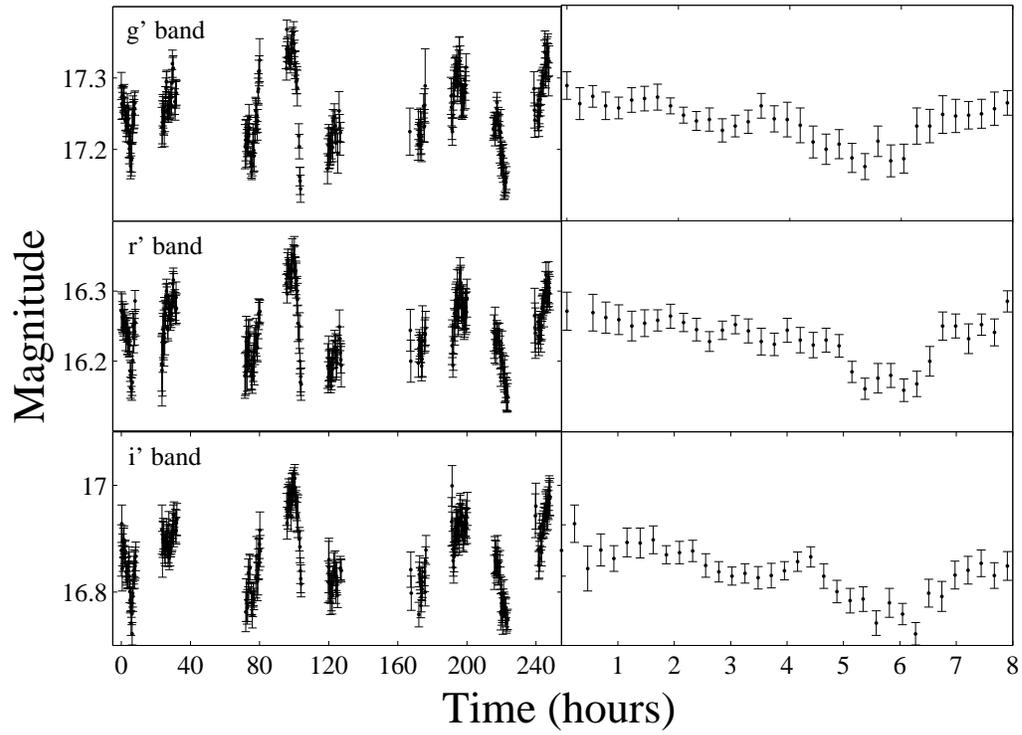}
    \caption{Light curves of NGC 4395 for g$'$, r$'$, and i$'$ filters. {\it left}: All nine nights. {\it right}: The first night only.  Time $=$ 0 is JD $=$ 2455616.3103. The data are listed in Table 1.}
\end{figure}

\begin{figure}[]
  \hskip -10.0cm
    \includegraphics[scale=0.7]{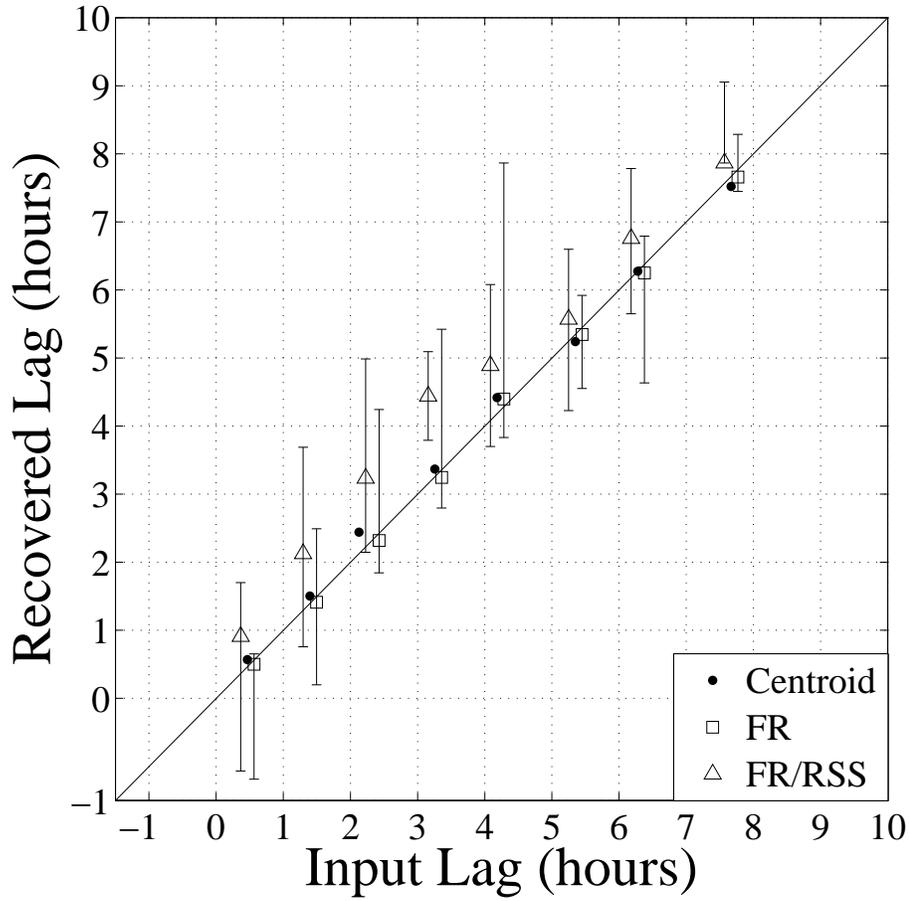}
  \caption{Recovered time lags for a series of simulated light curves shifted by some input time lag using the centroid method (black dots) on one hand, and a MC distribution of lags using the FR (open squares) and FR/RSS (open triangles) on the other.  The centroid and FR methods provide reliable lag estimates, while the FR/RSS usually overestimates the lags.}
\end{figure}

\begin{figure}[]
  \hskip 0cm
    \includegraphics[scale=0.8]{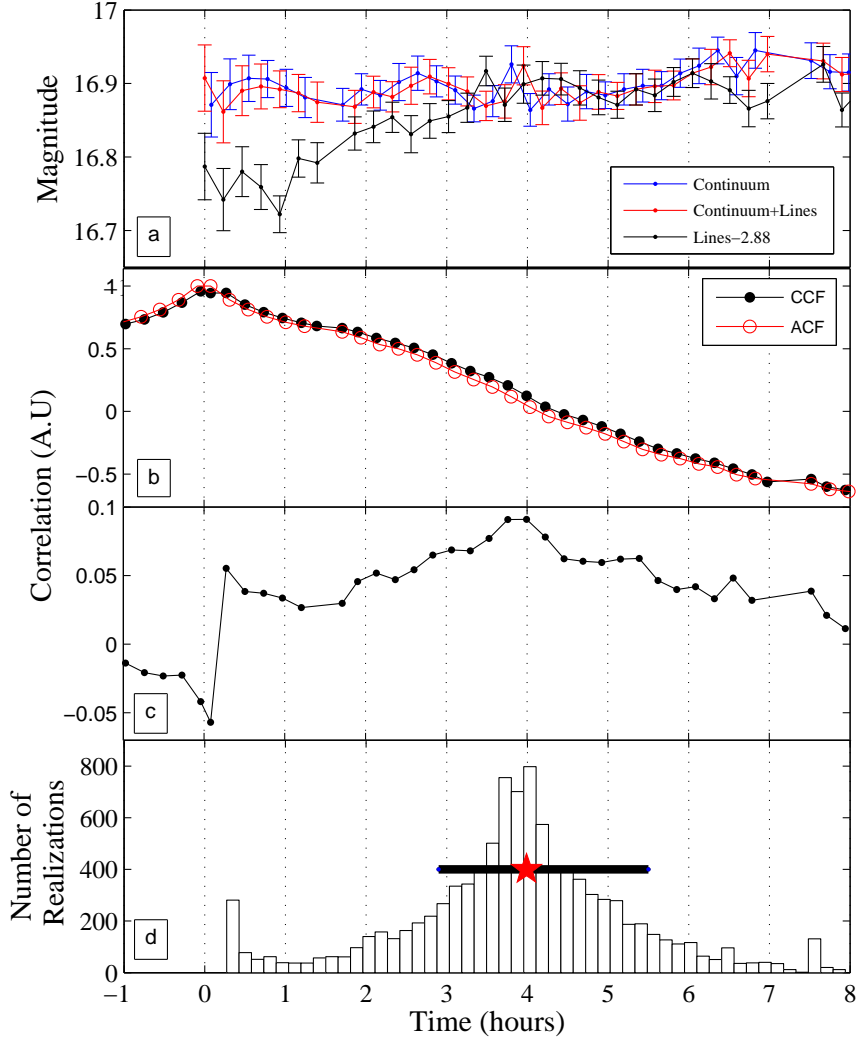}
    \caption{ {\it (a)}:  Simulated `continuum' light curve (LC), scaled `line' LC lagging the `continuum' by an arbitrary time of 4 hours (lowered by 2.88 magnitudes for plotting purposes), and the resulting `line$+$continuum' LC (we show only one night, but all nights are used in the simulation).  {\it (b)}:  CCF and ACF from the above light curves, and {\it (c)} the CCF$-$ACF difference which signifies a time lag of 4 hours.  {\it (d)}:  Histogram of 10,000 MC realizations of the time lag using the FR method producing a peak near the input lag and providing an estimate of the measurement error (4.0$^{+1.5}_{-1.1}$).}
\end{figure}

\begin{figure}[]
 \includegraphics[scale=0.7]{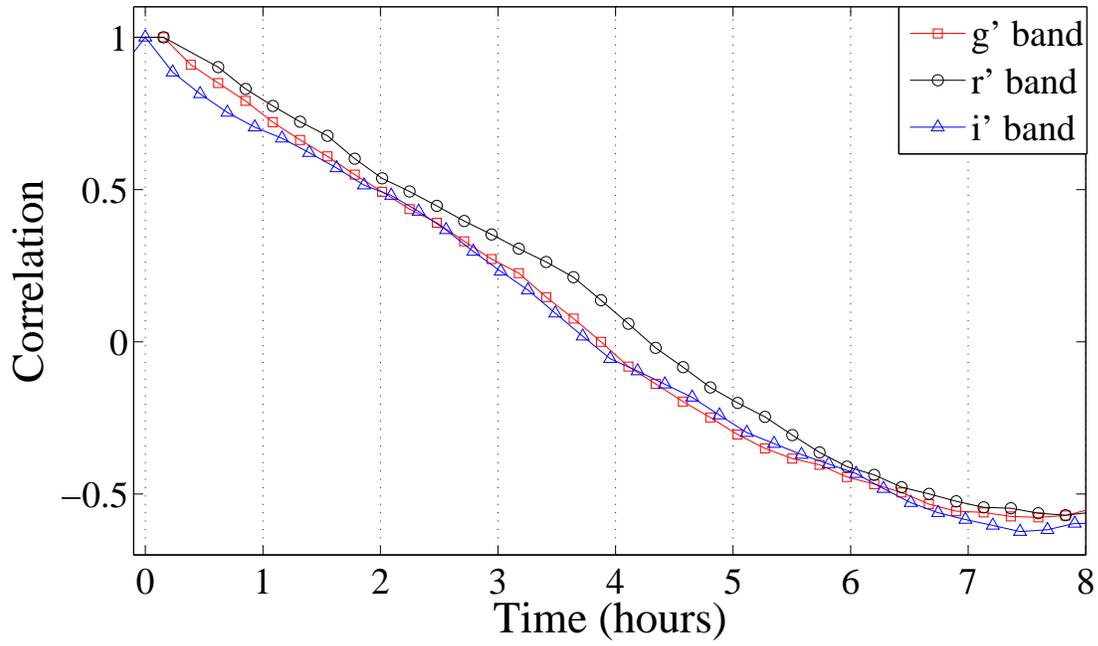}
 \caption{ACFs of all three filters.  A clear departure in the r$'$ band is seen at a time of about 4 hours compared to both the g$'$ and i$'$ bands, indicating perhaps an embedded BLR component.}
\end{figure}

\begin{figure}[]
  \begin{center}
    \includegraphics[scale=0.5]{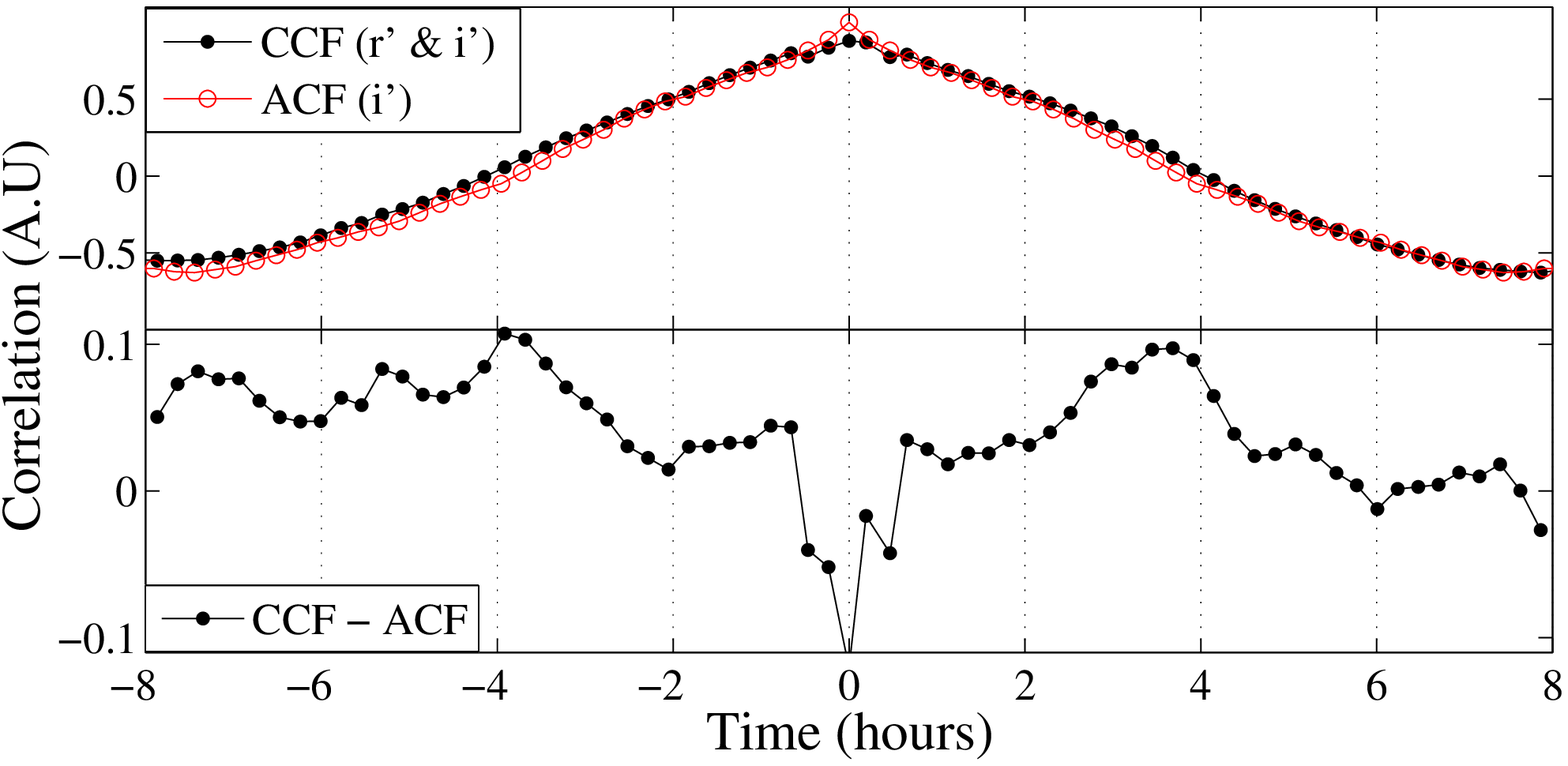}\\
    \includegraphics[scale=0.5]{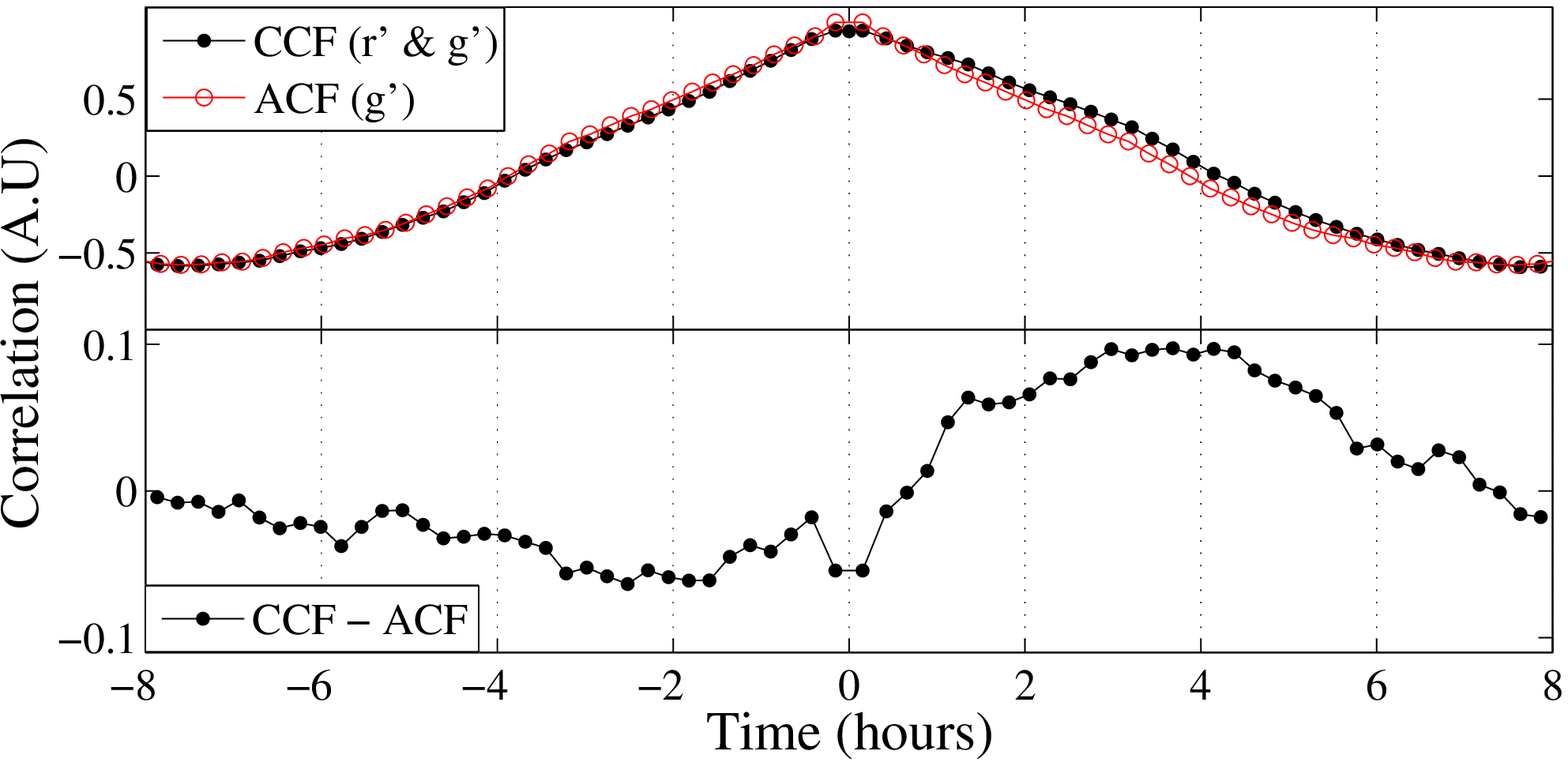}\\
    \includegraphics[scale=0.5]{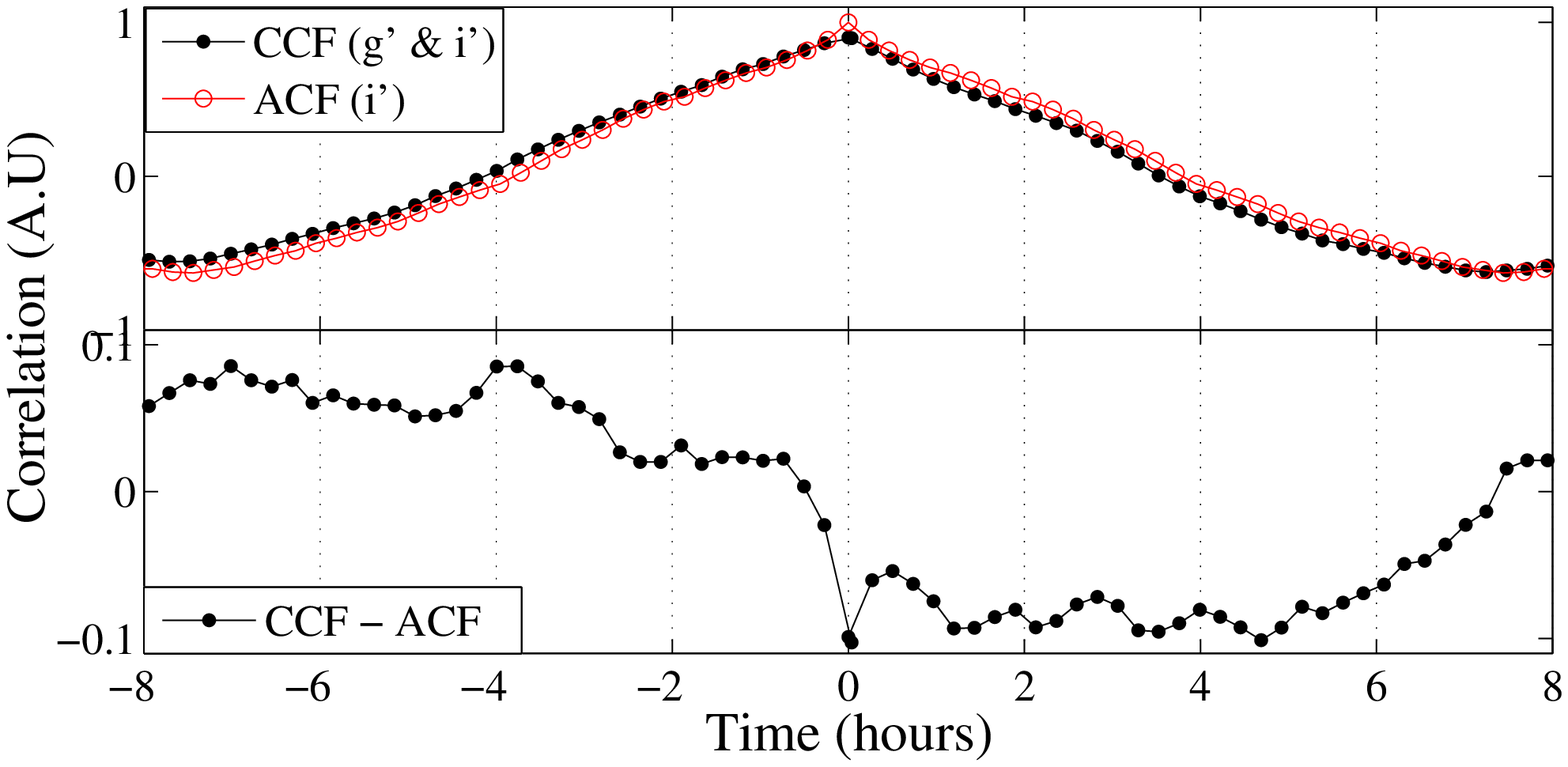}\\
    \end{center}
 \caption{CCF and ACF results for the three filter pairs. The lower panel in each plot shows the difference CCF-ACF, from which we compute the time lag.  Peaks can be seen in the two upper plots, indicating a time lag of about 3.5 hours.  No peak is apparent in the lower plot due to the weakness of the H$\beta$ line.  Features at negative times are discussed in the text.}
\end{figure}

\begin{figure}[]
\hskip 0.1cm
\includegraphics[scale=0.5]{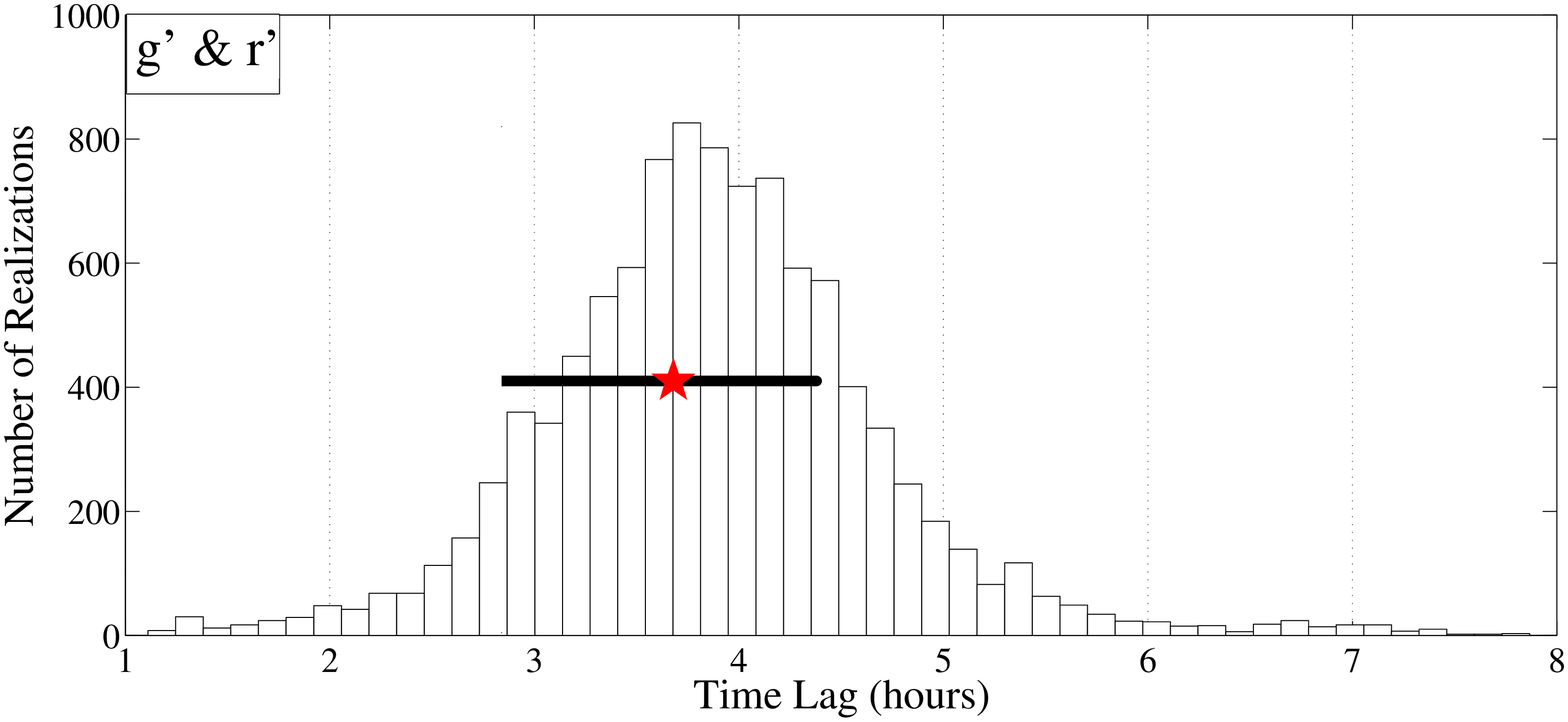} 
\includegraphics[scale=0.5]{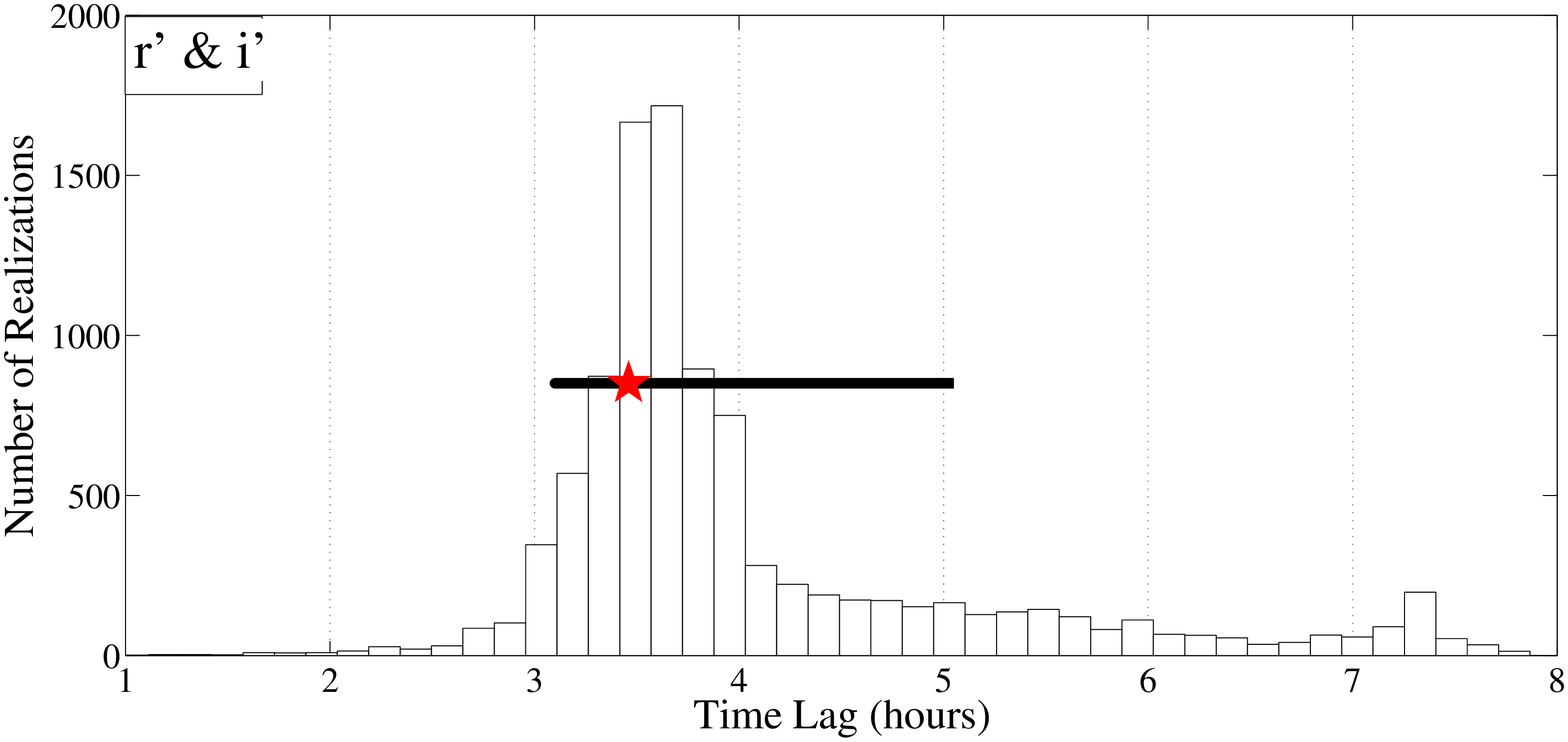}
  \caption{Distribution of time centroids obtained from 10,000 MC light curve realizations using the FR method for the filter pairs: r$'$ \&\ i$'$ (top) and r$'$ \&\ g$'$ (bottom).  The time lag as determined by the CCF-ACF difference in Figure 6 is shown as a star.  Horizontal bars indicate 68\%\ of the total realizations around the most probable time lag and represent its uncertainty range.}
  \label{fig:edge} 
\end{figure}

\begin{figure}[]
  \hskip -18.0cm
    \includegraphics[scale=0.8]{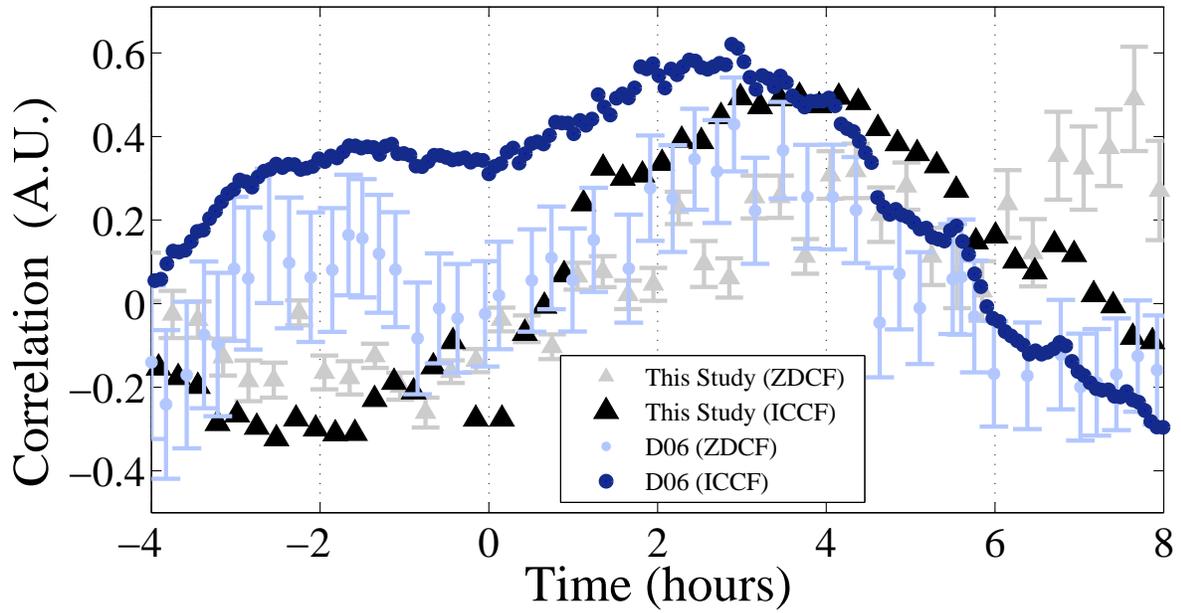}
  \vskip -6cm
  \caption{Results from spectroscopic RM of D06 compared to the Photometric RM results using the ICCF and ZDCF analysis. Similar time lags can be traced by all methods, as the peaks are all centered between 2-4 hours.}
\end{figure}

\end{document}